\documentstyle[aaspp4,epsfig,12pt]{article}
\def\paren#1{\left( #1 \right)}

\def\angle#1{\left\langle #1 \right\rangle}
\begin{document}
\title{Ultra Efficient Internal Shocks}
\author{Shiho Kobayashi$^1$ and Re'em Sari$^2$}
\affil{$^{1}$Department of Earth and Space Science, Osaka University,
Toyonaka, Osaka 560, Japan\newline
$^{2}$Theoretical Astrophysics 130-33, California Institute of Technology,
Pasadena, CA 91125, USA}

\begin{abstract}
 Gamma-ray bursts are believed to originate from internal shocks which
 arise in an irregular relativistic wind. The process has been thought
 to be inefficient, converting only a few percent of the kinetic
 energy into gamma-rays. We define ultra efficient internal shocks as
 those in which the fraction of emitted energy is larger than the
 fraction of energy given to the radiating electrons at each collision. 
 We show that such a scenario is possible and even plausible. In our
 model, colliding shells which do not emit all their internal energy
 are reflected from each other and it causes subsequent collisions,
 allowing more energy to be emitted.  As an example, we 
 obtain about $60\%$ overall efficiency even if the fraction of energy
 that goes to electrons is $\epsilon_e=0.1$ provided that the shells' Lorentz
 factor varies between 10 and $10^4$.
 The numerical temporal profile reflects well the activity of the source
 which ejects the shells, though numerous collisions take place in
 this model.
\end{abstract}
\keywords{ gamma rays: bursts;  shock waves; relativity}
\section{Introduction}

A widely accepted mechanism for producing a cosmological gamma-ray
burst (GRB) is the deceleration of relativistically expanding shells.
The kinetic energy of the shells is converted into internal energy
by relativistic shocks. These shocks can be due to collision with the
ambient medium (external shocks) or shocks inside the shell itself due
to nonuniform velocity (internal shocks). Electrons are heated by the
shocks, and the internal energy is radiated via synchrotron and IC
emission, with broken power law spectra (e.g. Sari, Piran and Narayan
1998).

Most bursts have a highly variable temporal profile with
variability time scale significantly shorter than the overall duration.  In
the external shocks scenario this variability is due to irregularity
in the surrounding material, but the efficiency is extremely low
(Fenimore, Madras \& Nayakshin 1996; Sari \& Piran 1997). Thus, GRBs
are believed to be produced by the other alternative: internal
shocks. The inner engine itself should be variable in this scenario,
the observed temporal profile follows very closely the operation of
the source (Kobayashi, Piran \& Sari 1997, KPS97 hereafter). 
Using a simple model, we
have estimated that the hydrodynamic efficiency of this process
(transforming kinetic energy to internal energy) is about $10\%$ .

Kumar (1999) argued that the conversion efficiency from the bulk
motion to gamma-ray is only $1\%$. His argument is based on three
points: (I) The hydrodynamic efficiency is, as mentioned above,
typically $10\%$. (II) It is only the electrons that are
radiating. Even in equipartition among protons, magnetic field and electrons,
the latter electrons only have a third of the internal energy. (III) The
amount of the radiated energy within the gamma-ray band is about a
third of the total radiated one. Combining these three factors give
the low efficiency of $\frac{1}{10}\times \frac{1}{3}\times
\frac{1}{3}=1\%$. 

Such a low efficiency results in sever energy demands on the source.
Moreover, it has problems explaining the energy ratio between the GRB
and its afterglow.
According to the internal-external shock model, the remaining kinetic
energy, which was not converted to radiation by internal shocks, is
radiated during the afterglow stage. External shock does not suffer
from problem (I), and the energy released in the afterglow should be
considerably higher than that in the GRBs. However, it seems to be
that the energy during the afterglow is only a tenth of that during
the GRB, rather than ten times larger (Frontera et. al. 2000, Kumar \&
Piran 2000, Freedman \& Waxman 2000). 
Even though the observational constraints are not
very good since most of the energy is released at very early radiative
stages where the afterglow was not observed, a factor of ten more
energy in the afterglow seems to be excluded.

A possible solution to these problems is to assume large angular fluctuation in
the shells (Kumar and Piran 1999). This model has clear predictions in
the form of afterglow variability whose amplitude decay in time. It
also predicts that the afterglow may be sometimes more energetic than
the GRBs.

In this paper, we suggest a simple alternative solution, which
overcomes the problems suggested by Kumar. We show that if
the distribution of the Lorentz factor is not uniform, but instead its
logarithm is distributed uniformly, then the typical ratio of Lorentz
factors between neighboring shells is considerably larger. Then, the
hydrodynamic efficiency can be close to hundred percents, even for a
reasonable spread of Lorentz factors (Similar calculation was recently
done by Beloborodov 2000). The main point of this paper is the
possibility of ``ultra-efficient'' internal shocks. We define
ultra-efficient internal shocks as a scenario in which the 
emitted fraction of kinetic energy is larger than $\epsilon_{e}$, 
the fraction of internal energy that is going into electrons (and then
radiated) at each collision. We will show that such a scenario is
possible and even reasonable. 

\section{``Ultra Efficient'' Internal Shocks}

Internal shocks could occur within a variable relativistic wind 
produced by a highly variable source. We represent the irregular 
wind by a succession of relativistic shells with a random distribution
of Lorentz factors in a similar manner as in KPS97. Beloborodov (2000)
has shown that the internal shocks can convert most of the kinetic
energy to internal energy if the fluctuation of the initial Lorentz 
factors $A^2=(\angle{\gamma^2}-\angle{\gamma}^2)/\angle{\gamma}^2$ 
is large. Though it
is maximally 1/3 if the initial Lorentz factors take random
values between $\gamma_{min}$ and $\gamma_{max}$, it is not limited
if the distribution is uniform in logarithmic space between 
$\log\gamma_{min}$ and $\log\gamma_{max}$. 

\begin{figure}[b!] 
 \centerline{\epsfig{file=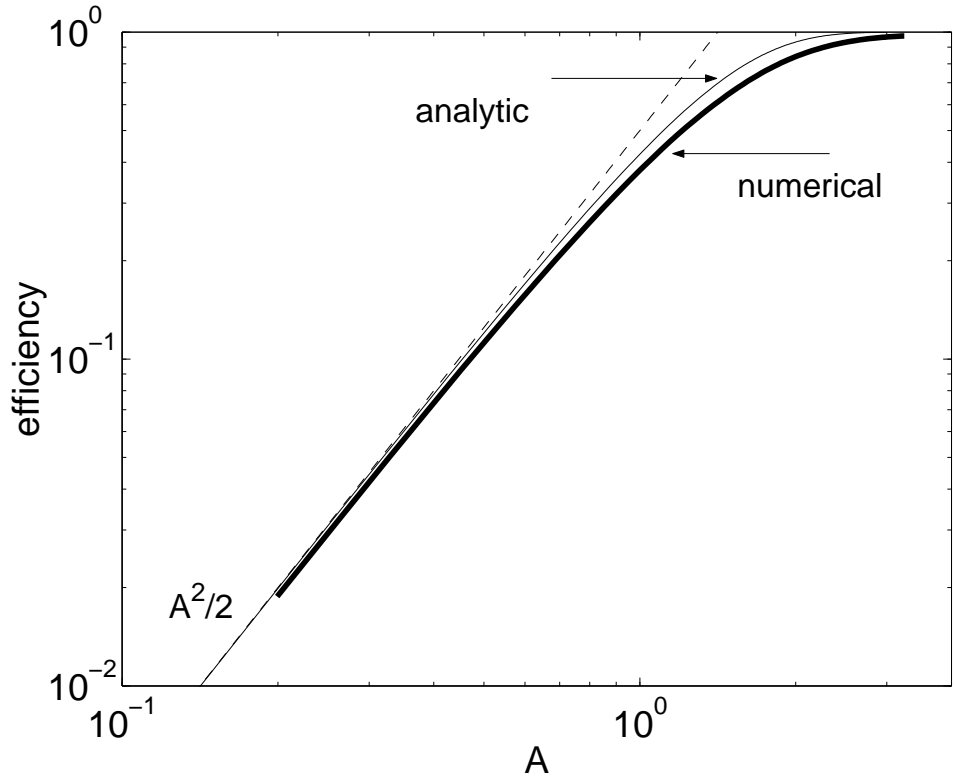,width=5in}} \vspace{10pt}
 \caption{Fully radiative case ($\epsilon_e=1$): efficiency vs $A$~.  
thick solid (numerical simulation: average of 100 random 
simulations with 100 equal mass shells), thin solid (analytic
estimate), dashed ($\angle{\epsilon}=A^2/2$)
} 
 \label{fig:gmm}
 \end{figure}

The efficiency can be estimated by an equation similar to
equation 19 in KPS97. The most
efficient case is that the masses of the shells are taken equal, and
that the efficiency is given by a simple form:
\begin{equation}
\angle{\epsilon}\sim 1-a^{1/2}\log a/(a-1)
\end{equation}
where $a=\gamma_{max}/\gamma_{min}$. The efficiency is plotted as a 
function of the fluctuation $A^2=-1+(a+1)\log a/2(a-1)$ in figure
\ref{fig:gmm}. This analytic estimate fits the result of our numerical
simulation, and has the asymptotic form $\angle{\epsilon}=A^2/2$
which Beloborodov estimated for a small fluctuation. Our expression
above generalizes Beloborodov estimate and gives reasonable estimate of
the efficiency also for large fluctuations. 

Though we have seen that a large hydrodynamic efficiency is possible if
the fluctuation of the initial Lorentz factors is large, it is
not reasonable that all the internal energy is emitted after each
collision, since electrons do not have most of the internal energy.
Defining $\epsilon_e$ as the fraction of energy given to electrons
we expect $\epsilon_e<1$. Even at equipartition with protons
$\epsilon_e=1/2$. Under these circumstances the total emitted energy in
Beloborodov model is still limited by $\epsilon_e$.

However, if $\epsilon_e<1$ the merger produced by a collision is
expected to stay hot after the emission. As a result, the merger will
spread to transform the remaining internal energy back to the kinetic
energy (Kumar 1999). A simplified description of this process is to
assume that the two shells reflect with a smaller relative velocity. The
difference of the kinetic energy before and after the collision is the
emitted internal energy. The reflecting shell will collide into the
other neighbor shell. Since in this way a large amount of collisions 
are caused, the overall efficiency from the kinetic energy to the
radiation could be larger than $\epsilon_e$. This is the key
ingredient of our model. Before going on with this model, we present
some hydrodynamic simulations which shows that these simplified
assumptions are reasonable.

\section{Hydrodynamic Simulation}
To estimate the conversion efficiency of the internal shocks process, it
is important to understand how high velocity shells interact with 
slower ones and dissipate the the kinetic energy. We consider here a
collision of two equal mass shells with very different Lorentz factors,
$\gamma_r/\gamma_s \gg 1$. The rapid and the slower shells are denoted
by the subscriptions $r$ and $s$ respectively. We assume that the widths
of the shells are comparable in the ISM rest frame. This is a reasonable
assumption since in this frame, the width is given directly by the
``inner engine''. Even so, the width of the rapid shell $\hat{l}_r$ is
much larger when viewed from the rest frame of the rapid shell
$\hat{l}_r/\hat{l}_s \sim (\gamma_r/\gamma_s)^2/2$. Then, the dense
``slower'' shell with Lorentz factor 
$\hat{\gamma}_s \sim \gamma_r/2\gamma_s$ collides into the low
density ``rapid'' shell in this frame,
$\rho_s/\rho_r \sim \gamma_r/\gamma_s$. This is a planar analog
of the evolution of a relativistic fireball (Sari \& Piran 1995, 
Kobayashi, Piran and Sari 1999). When it begins to interact with
the surrounding material, two shocks are formed: a forward shock
propagating into the rapid shell and a reverse shock propagating into
the slower one. 
After the reverse shock crosses the slower shell, the profile of the 
shocked rapid shell material approaches to that of its 
fireball analog: the ``blast wave''
which sweeps and collects the surrounding material. Once the shock wave
crosses the rapid shell, the hydrodynamical structure is as follows. 
There is a shocked rapid shell, the analog of the ``blast
wave'' and the shocked slow shell, which cooled down and is the 
analog of the adiabatically cooling ``fireball ejecta''.  The structure
of such a system in the fireball case was studied by Sari and Piran 1999
and Kobayashi and Sari 2000.

Since $\hat{\gamma}_s^2 \gg \rho_s/\rho_r$, the slow shell is 
considerably decelerated by the relativistic reverse shock down 
to $\sim (\gamma_r/\gamma_s)^{3/4}$ and heated to a relativistic
temperature at the crossing time $t_s \sim (\gamma_r/\gamma_s)^{3/2}
\hat{l}_s/c $. Then, it cools adiabatically and follows a planar version
of the Blandford Mckee (1976) solution (Sari 2000), in which a given
fluid element evolves with a bulk Lorentz factor of $\gamma \propto
l^{-3/2}$. Therefore, the motion becomes Newtonian
$\gamma \sim (\gamma_r/\gamma_s)^{3/4} (t_r/t_s)^{-3/2} \sim 1$ 
when the forward shock crosses the rapid shell at $t_r=\hat{l}_r/c$.
On the other hand, since the forward shock itself evolves as 
$\gamma \propto l^{-1/2}$, it slows down to 
$\gamma \sim (\gamma_r/\gamma_s)^{1/2}/2$ at the crossing time $t_r$. 

We have developed a relativistic code with an exact Riemann solver to
solve relativistic hydrodynamics problems (Kobayashi, Piran \& Sari
1999). Using this code, we numerically study the collision of two  
equal mass slabs with $\gamma_s=10$, $\gamma_r=10^3$ and the same width 
in the ISM frame. The initial condition in the rapid shell comoving
frame is  
\begin{center}
\begin{tabular}{ll}
  $\hat{\gamma}_r=1$    &  $\hat{\gamma}_s \sim 50$ \\
  $\rho_r=  1$          &  $\rho_s   =100$          \\
  $\hat{l}_r \sim 5000$ &  $\hat{l}_s=1$            \\
\end{tabular}
\end{center}
The mass density outside the slabs and the homogeneous pressure are
$\rho=10^{-8}$ and $p=10^{-10}$. The mass density and the pressure are 
measured in the comoving frame of each fluid. Our adiabatic simulations
represent a case in which $\epsilon_e$ is very small.    
In this simulation, we used the unit of $c=1$.

\begin{figure}[b!] 
 \centerline{\epsfig{file=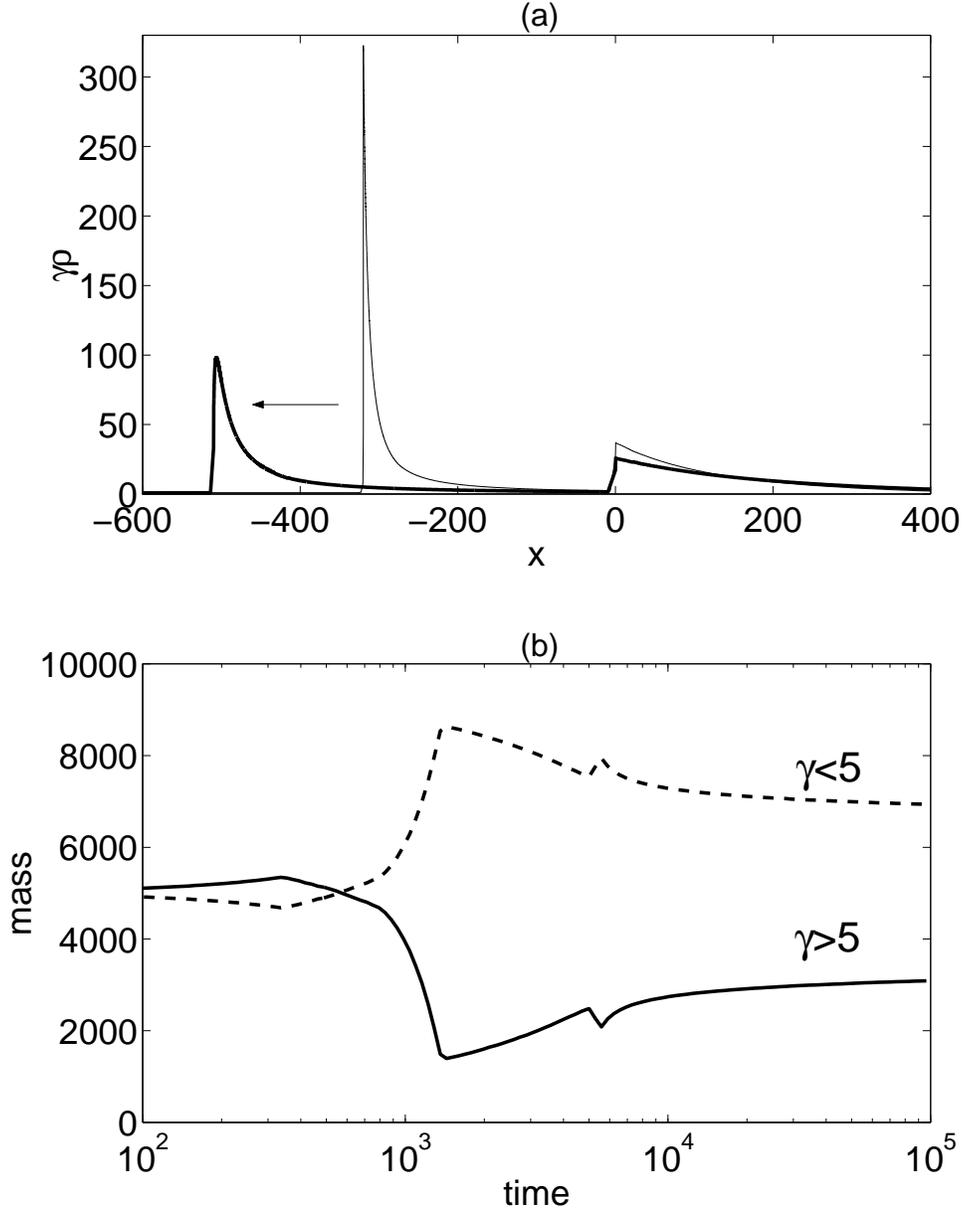,width=5in}} \vspace{10pt}
 \caption{
(a) Lorentz factor $\gamma$ vs distance x from the
contact surface at $t=5000$ (thin) and $6000$ (thick).
The Lorentz factor $\gamma$ and position $x$ are measured in the
rapid shell frame, and the density $\rho$ is the the fluid local frame.
(b) mass fractions: $\gamma>5$ (solid) and $\gamma<5$ (dashed) 
 } 
 \label{fig:2shells}
 \end{figure}

To compare the analytical estimates with the numerical simulation we
define the effective Lorentz factor of each shell as $\langle \gamma
\rangle=\int m\gamma dl/\int m dl$ with the effective mass $m=\gamma
\{\rho+(3+\beta^2)p \}$.  The numerical simulation then gives $\langle
\hat{\gamma}_r \rangle \sim 5.3$ and $\langle \hat{\gamma}_s \rangle
\sim 1.6$ at the crossing time, which are in agreement with the 
analytical estimates. The thin line in figure \ref{fig:profile} shows
the numerical density profile at this time.
After the forward shock crosses the rapid
shell, a rarefaction wave begins to propagate into the rapid shell to
transform the internal energy to the kinetic, then the rapid is
accelerated to $\gamma \sim \gamma_r/2\gamma_s$.
The center of mass moves with a Lorentz factor $\sim
\paren{\gamma_r/\gamma_s}/2\sim5$ in the rapid shell comoving frame. At 
the end of the simulation $t=10^5$,  about forty percent of the rapid
shell material is slower than the center of mass, and goes with
the slower shell (see fig \ref{fig:2shells}b). 

\section{Two Shell Collision}

A collision of two shells is the elementary process in our model.
A rapid shell catches up a slower one and the two merge to temporarily 
form a 
single one (denoted by the subscript m). Using conservation of energy
and momentum, the Lorentz factor of the merged shell $\gamma_m$ and the
internal energy $E_{int}$ produced by the collision are given by 
\begin{equation}
\gamma_m\sim 
\sqrt{(m_r\gamma_r+m_s\gamma_s)/(m_r/\gamma_r+m_s/\gamma_s)}~,
~E_{int}= m_r(\gamma_r-\gamma_m)+m_s(\gamma_s-\gamma_m).
\end{equation}

After a fraction $\epsilon_e$ of the internal energy is emitted
isotropically in the local frame of merged shell (center of mass frame), 
the shells will spread to transform the remaining internal energy back
to kinetic energy. If the widths are the same in center of mass frame,
each mass is conserved before and after the collision. However, as we 
have seen, some fraction of the rapid shell material goes together 
with the slower one after the collision in general. We parameterize the
mass splitting as 
$m_r^\prime=(1-\delta)m_r$ and $m_s^\prime=m_s+\delta ~m_r$.
We later show that the total efficiency is not sensitive to 
$\delta$.

The Lorentz factor of the reflected shells in center of mass frame 
are given by   
\begin{equation}
\bar{\Gamma}_r=(M^2+m_r^{\prime ~2}-m_s^{\prime ~2})/2m_r^\prime M, \ \ 
\bar{\Gamma}_s=(M^2+m_s^{\prime ~2}-m_r^{\prime ~2})/2m_s^\prime M.
\label{twoshellsgamma}
\end{equation}
where $M=(m_r\gamma_r+m_s\gamma_s-\epsilon_eE_{int}/c^2)/\gamma_m$.
The Lorentz factors in laboratory frame are  
\begin{equation}
\Gamma_r=\bar{\Gamma}_r\gamma_m-\sqrt{(\bar{\Gamma}_r^2-1)(\gamma_m^2-1)}, 
\ \ \ \ 
\Gamma_s=\bar{\Gamma}_s\gamma_m+\sqrt{(\bar{\Gamma}_s^2-1)(\gamma_m^2-1)}.
\label{eq:tolab}
\end{equation}
The shells are once compressed by shocks, but these spread when
reflecting. For simplicity we assume that the width of the shells, 
$l_i$, is constant. 

\section{Multiple Shell Collision}

We consider a wind consisting of $N$ shells. Each shell is 
characterized by the four variables: $\gamma_i, m_i, l_i$ and $R_i$. 
We assume that the initial Lorentz factor of each shell is distributed 
uniformly in logarithmic space between $\log \gamma_{min}$ and 
$\log \gamma_{max}$. The initial masses are assumed to be correlated
with the Lorentz factors as $m \propto \gamma^\eta$.  For $\eta =-1$ and
$0$ the shells initially have equal energy and equal mass respectively,
and for $\eta =1$ the shells initially have equal density under an
assumption of the equal shell width. We assume a constant value $l$ for
the initial widths and the initial separations between the shells. Then,
the initial position of the shells are $R_i=2(i-1)l$.  

The evolution of the system in time is basically same as in KPS97,
but equation \ref{eq:tolab} is used to calculate the Lorentz factor
of the reflecting shells for the next time step. We follow the evolution
of shells until there are no more collisions, i.e. until the shells
are ordered with increasing value of the Lorentz factors.

The conversion efficiency from the kinetic energy of the shells to 
radiation can be calculated by using the initial and final kinetic
energy as $\angle{\epsilon}=1-\Sigma m_i^{(f)}\gamma_i^{(f)}/
\Sigma m_i^{(i)}\gamma_i^{(i)}$ where the superscripts $(f)$ and $(i)$ 
represent the initial and final values respectively. It depends on the
model parameters $\{\gamma_{max}/\gamma_{min}$, $N$, $\eta,
\epsilon_e, \delta \}$ and on the specific realization: the set of
random Lorentz factors assigned to each shell. For each choice of the
parameters of the model, we have evaluated the efficiency for 100
realizations. The mean efficiency and its standard deviation are listed
in Table 1 for the conserved equal mass case ($\eta=\delta=0$). 
\begin{center}
\tabcolsep=3pt \footnotesize
Table.1 \\
\vspace{0.5cm}
\begin{tabular}{|r|c|c|c|c|} \hline
$N$ & $\gamma_{min}$& $\gamma_{max}$ & efficiency [\%] $(\epsilon_e=0.1)$
& efficiency [\%] $(\epsilon_e=0.5)$ \\ \hline
30    & $10^2$& $10^3$ &$ 9.2\pm2.3$ & $ 16.2\pm3.0$ \\ 
30    & $10  $& $10^4$ &$40.0\pm9.2$ & $ 67.5\pm9.3$ \\ \hline
$10^2$& $10^2$& $10^3$ &$15.1\pm1.5$ & $ 17.7\pm1.5$ \\ 
$10^2$&  10   & $10^4$ &$62.9\pm4.8$ & $ 72.4\pm4.3$ \\ \hline
\end{tabular}
\end{center}

\begin{figure}[b!] 
 \centerline{\epsfig{file=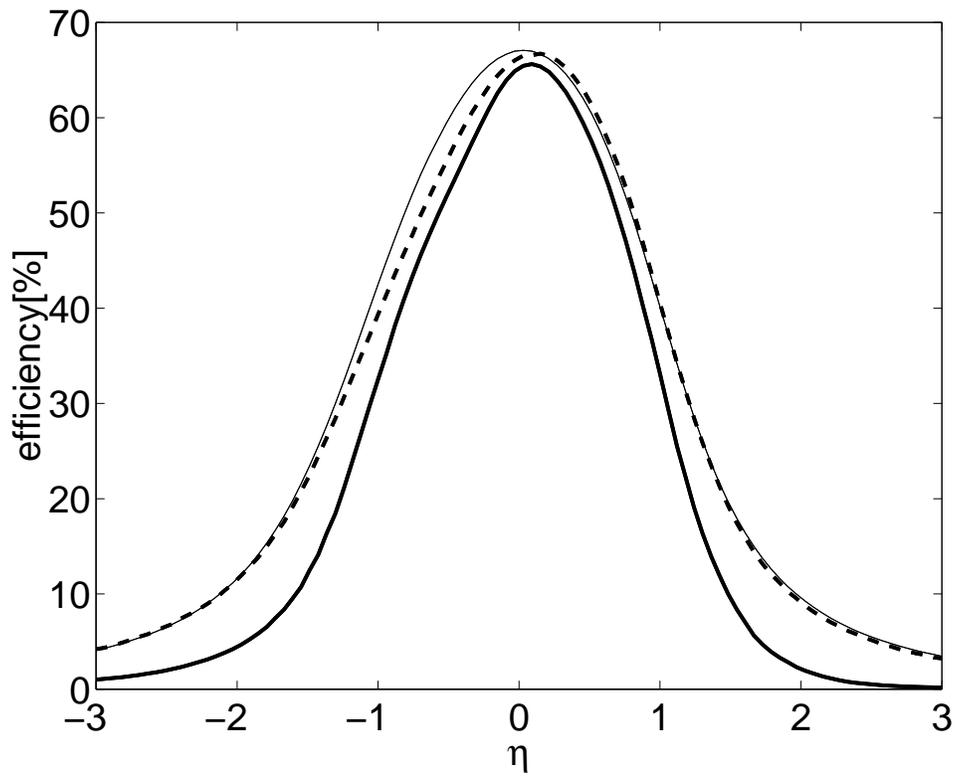,width=5in}} \vspace{10pt}
 \caption{ The efficiency is plotted as a function of 
$\eta$ for different $\delta$.
$\delta=0$ (thick solid), 0.4 (dashed) or 0.8 (thin).
$N=30, \gamma_{min}=10$, $\gamma_{max}=10^4$ and $\epsilon_e=0.5$.
} 
 \label{fig:eta}
 \end{figure}

The efficiency approaches an asymptotic value as the Lorentz factor
ratio $\gamma_{max}/\gamma_{min}$ and the number of the shells $N$
increase. The asymptotic value depends on $\eta$, $\epsilon_e$ and 
$\delta$. The efficiency is plotted in fig \ref{fig:eta} as a 
function of $\eta$ for $\delta=0, 0.4$ or 0.8. In the range 
$-1 \le \eta \le 1$, it is not very sensitive to $\delta$ and 
peaks around $\eta\sim0$, i.e. the most efficient case
is the equal mass case. 
 
\begin{figure}[b!] 
 \centerline{\epsfig{file=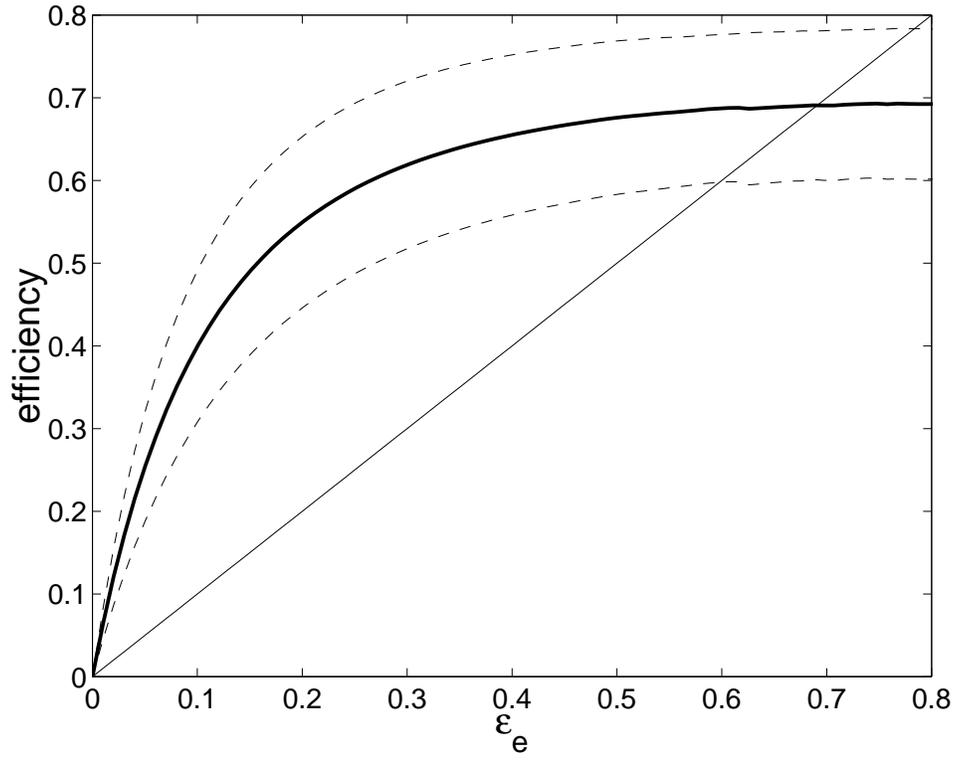,width=5in}} \vspace{10pt}
 \caption{Efficiency vs $\epsilon_e$ with $1 \sigma$ error bars
of 100 random simulations.
$N=30$, $\eta=\delta=0$, $\gamma_{min}=10$ and $\gamma_{max}=10^4$.
} 
 \label{fig:eff_eps}
 \end{figure}

The efficiency is plotted as a function of $\epsilon_e$ for the 
conserved equal mass case ($\eta=\delta=0$) 
in Figure \ref{fig:eff_eps}. It is interesting
that the efficiency can be larger than $\epsilon_e$ since the energy
released in GRB can be much larger than that in the afterglow.  For
instance, consider the case of $N=100$, $\gamma_{min}=10$,
$\gamma_{max}=10^4$, $\eta=\delta=0$ and  $\epsilon_e=0.1$, the internal
shocks can convert about sixty percent of the kinetic energy to the
radiation. The reminder is converted to the thermal energy by external
shock. However, only the fraction $\epsilon_e=0.1$ of that is emitted as
the  afterglow
\footnote{In fact, Sari 1997 have shown that also during the afterglow
the fraction of energy that can be emitted may exceed
$\epsilon_e$. However, this will be spread over many decades of
time.}, 
the efficiency by the external shock is only four percent.
The GRB to the afterglow ratio for these parameters is about 15!.

\begin{figure}[b!] 
 \centerline{\epsfig{file=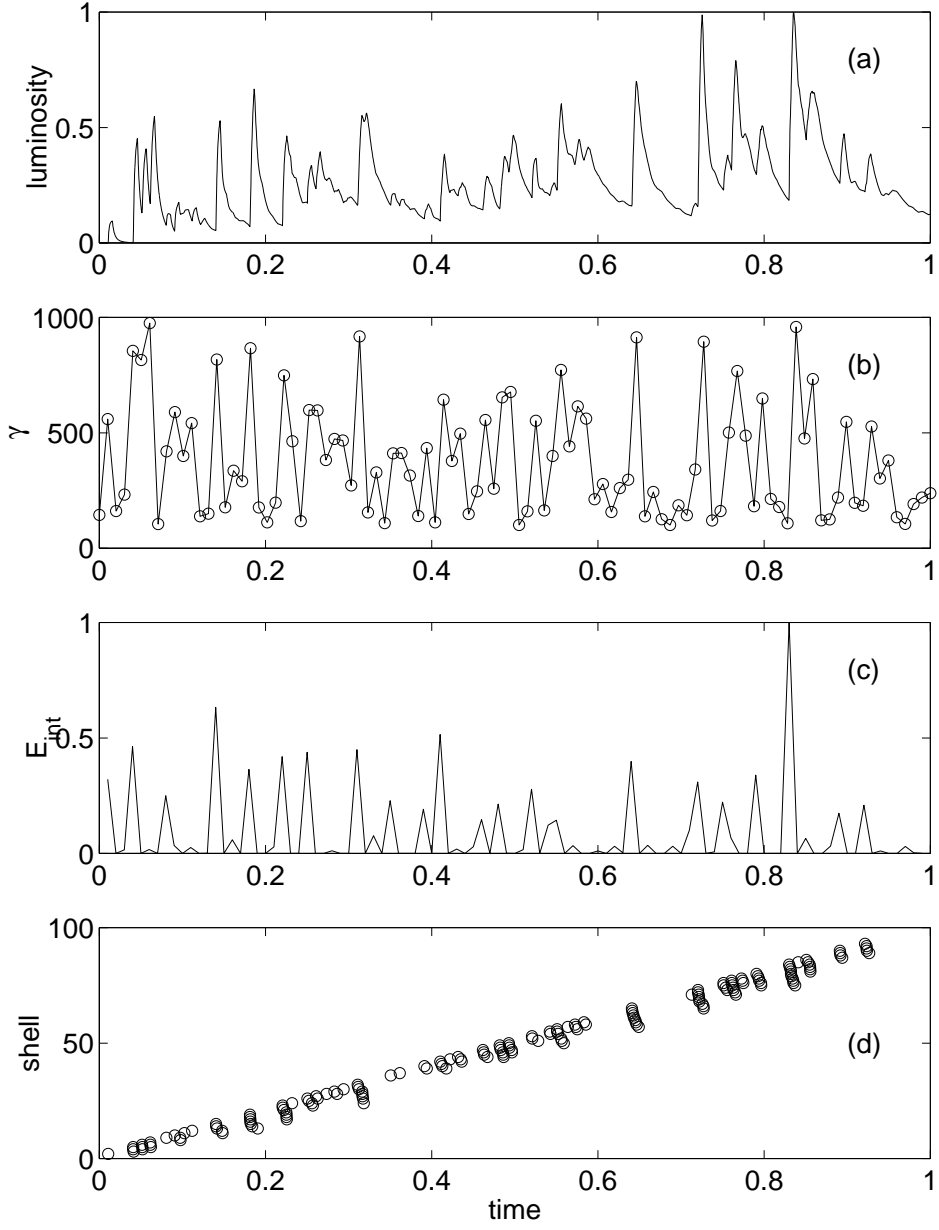,width=5in}} \vspace{10pt}
 \caption{%
Temporal structure.
(a) Numerical temporal structure. 
(b) Lorentz factor of a shell at the ejection vs ejection time
(c) $E_{int}$ between neighbor boring shells vs ejection time
(d) Index of a shell
vs observed time of the radiation produced in that shell.
$N=100, \gamma_{min}=10^2, \gamma_{max}=10^3, \eta=\delta=0$
and $\epsilon_e=0.1$.
} 
 \label{fig:profile2}
 \end{figure}

Figure \ref{fig:profile2}a shows the resulting temporal structure
in the equal mass case. It is a superposition of pulses from the
elementary two shell collisions. Though numerous collisions take place
during the evolution, the number of peaks in the profile is of order of
$N$.  Since all peaks have widths of the same order of magnitude, the
different amplitude of the peaks originates mainly from the difference
in the internal energy produced by the collisions.
In Figure \ref{fig:profile2}b, we plot the initial Lorentz factor as a 
function of time when shells were emitted by the source. We evaluate
$E_{int}$ for all pairs of the shells if inner shell is faster. We
assign $E_{int}$ to the ejection time of the inner shell (see Figure
\ref{fig:profile2}c). It resembles well the temporal profile in
\ref{fig:profile2}a. Therefore, despite the complicated nature of the
two shells collisions, the observed burst closely follows the inner
engine temporal profile.  

Neighboring shells collide on a time scale of $2\gamma^2 l/c$.  The
matter is moving toward the observer, the resulting observer time
scale is $l/c$. On the other hand, the difference in observer time due
to the location of the given shell within the wind is of order $Nl/c$.
Then, we observe the pulses arising from the collisions according to
their positions inside the wind. In Figure \ref{fig:profile2}d we plot
the shell's index against the time when a radiation from the shell is
observed. Although the light curve in figure \ref{fig:profile2}a is
the superposition of all pulses, in figure \ref{fig:profile2}d we
plot only the pulses which are higher than a tenth of the highest. 
We can see a clear correlation between the index and the time, the
temporal profile reflects the activity of the source. 

The deviation from the correlation, e.g. some groups of circles in a
line from left top to right bottom is due to that a rapid inner shell
collides into the outer neighbor, and the boosted shell further collides
into the outer. Since the masses of the shells are equal, the Lorentz
factors of the inner and the outer shells are just switched at each
collision if the radiation is negligible. Then, if an initial Lorentz
factor of a shell is peculiarly high, the index of the shell which
has the high Lorentz factor propagates outwards. In this sequence with
the radiation loss, the pulse from the collisions damps quickly, 
then, the overall correlation is not destroyed.

\begin{figure}[b!] 
 \centerline{\epsfig{file=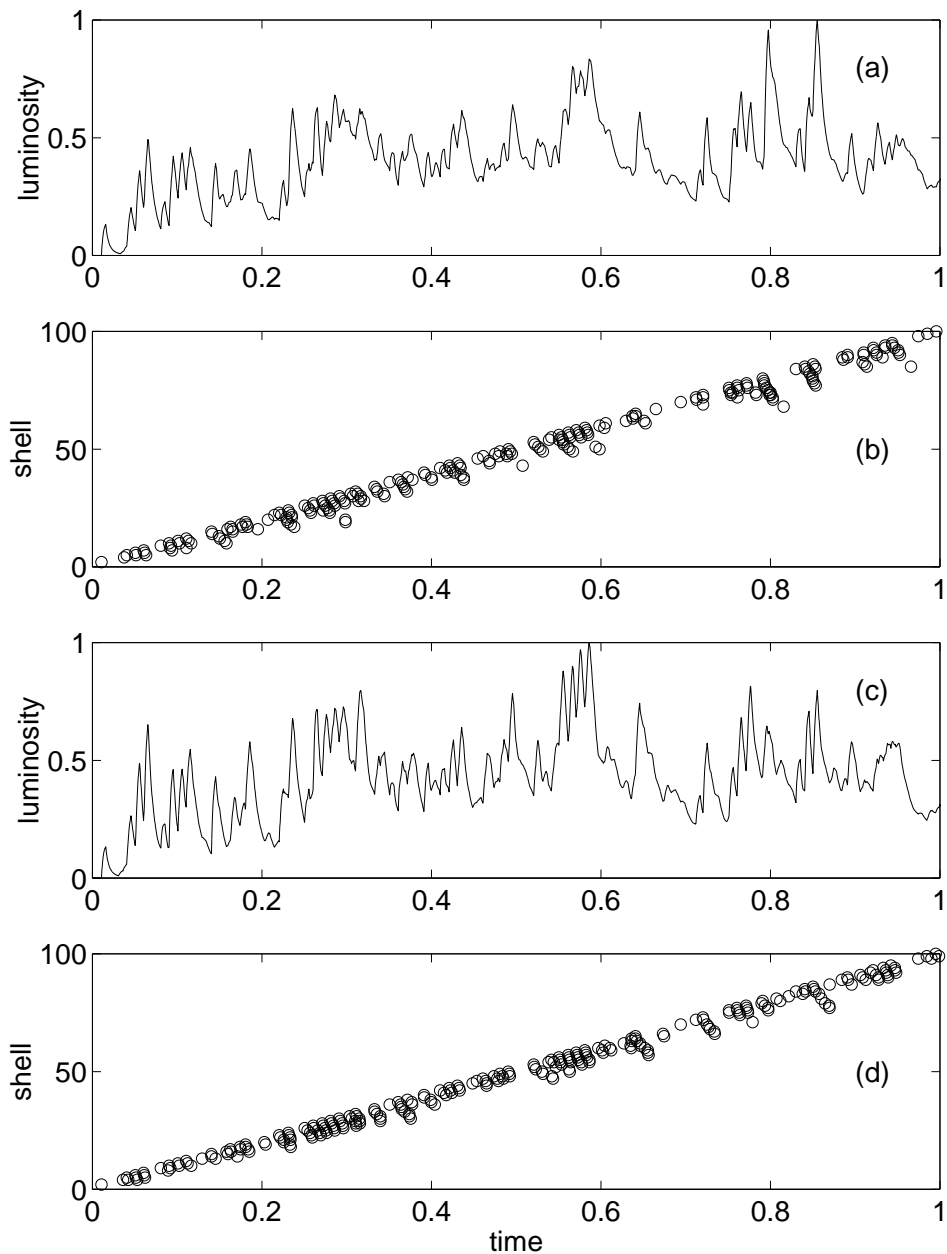,width=5in}} \vspace{10pt}
 \caption{
Temporal structure and correlation between the index of shell  
and the radiation observed time. The merger at each collision 
is assumed to split to the original masses  (a and b) or 
to the modified masses with $\delta=0.4$ (c and d).
$N=100, \eta=-1, \gamma_{min}=10^2$ and
$\gamma_{max}=10^3$ and  $\epsilon_e=0.1$.
} 
 \label{fig:profile}
 \end{figure}

The temporal structures for equal energy cases ($\eta=-1$) are plotted
in figure \ref{fig:profile}. The merger at each collision is assumed to
split to the original masses (figure \ref{fig:profile}a) or to the
modified masses $\delta=0.4$ (figure \ref{fig:profile}c). The initial
distributions of $\gamma$ are the same with that of figure
\ref{fig:profile2}. For $\eta=-1$, $E_{int}$ takes almost the same value
for most of the collisions (KPS97), so most of the peaks have comparable
amplitudes. The profiles still reflect the activity of the source
well. 

\section{Conclusions}

We have shown that the conversion efficiency from kinetic energy of
relativistic shells to radiation can be close to hundred percent
if the source produces shells with comparable masses with different 
Lorentz factor, especially when the logarithms of the Lorentz factors
 are distributed uniformly (see also Beloborodov 2000). It had been
assume in our previous work that the Lorentz factors themselves were
distributed uniformly, and the efficiency was less than forty
percent. With this distribution, the most efficient case is that the
source produces shells with comparable energy, instead of comparable
masses (KPS97). 

However, this high efficiency is achieved assuming that all the
internal energy is emitted at each collision, this is not reasonable
since only electrons radiative effectively, and these do not have all
the internal energy.  After the electrons radiate, large amount of the
internal energy remains in protons. Using a hydrodynamic simulation, it
has been shown that the hot merger produced by a collision spreads to
transform the remaining internal energy back to the kinetic one. A
simplified description of this process is to assume that the shells
reflect each other with a smaller relative velocity, after the
collision. The difference of the kinetic energy gives the radiated
internal energy.  

Since the reflecting shell collide into the outer neighbor shell, the
index of the shell which has a high Lorentz factor propagates outerwards
until the high value decays by the radiation loss. The shell itself
which has initially a high Lorentz factor might not go through many
collisions, but its high ``kinetic energy'' does. Therefore, the
internal shock process is very efficient even if the fraction of
internal energy emitted at each collision is small. Previously, the
efficiency in the case of $\epsilon_e <1$ had been estimated as
smaller by a factor of $\epsilon_e$ than that in the corresponding
fully radiative case. Our ultra efficient internal shocks scenario shows
this to be a significant underestimate. Though the efficiency that we
have estimated is bolometric efficiency, the efficiency from the kinetic
energy of the shells to gamma-ray radiation is also high if the faction
of the energy radiated in the BATSE band is not very small.

Numerous collisions happened in our ultra-efficient internal shocks
model, it made the peak width wider than in the previous internal
shocks model. However, the number of main peaks is still almost the same
as the number of shells that the source emitted.  There is a strong
correlation between the time at which we observe a pulse and the
emission time of the corresponding shell from the source. This
correlation persists even for a small $\epsilon_e$ case where larger
number of collision happen. The temporal structure reproduces the
activity of the source.  

We have shown that the efficiency of the internal shock process is not
limited by $\epsilon_e$, while that of the external shock is so. If a
fraction $\zeta$ of kinetic energy of an explosion is converted to
the radiation by the internal shocks, all the remaining one is
converted to the thermal by external shock in the afterglow stage,  
and a fraction $\epsilon_e$ of the thermal is emitted. Then, we can
roughly estimate the ratio between the energy released in afterglow and
that in GRB as  $\sim \epsilon_e (1-\zeta)/\zeta$. Assuming
$\epsilon_e=0.1$, the ratio is $1/10$ for $\zeta=0.5$ and decrease as
$\zeta$ increase. If the efficiency of internal shocks is indeed very
large, the luminosities of GRB and the afterglow are expected
to be anticorrelated. 

As a general argument in internal shock model, 
the luminosity distribution of multi peak bursts would be narrower than
that of bursts with only few peaks (Kumar \& Piran 2000),
since the number of the peaks is
almost the number of the shells $N$ as we have shown. We verified that
the dispersion of the efficiency is proportional to $1/\sqrt{N}$ in our
numerical model. 

S.K. acknowledges support from the Japan Society for the Promotion of
Science. R.S. acknowledges support from the Sherman Fairchild
foundation.

\noindent {\bf References}\newline
Beloborodov.A. 2000, ApJL, 539, L25.\newline
Fenimore,E.E., Madras,C.D. \& Nayakshin,S. 1996,ApJ,437,998.\newline
Freedman,D.L. \& Waxman,E. 1999, 
              Submitted to ApJ, astro-ph/9912214.\newline
Frontera F., et. al., 2000, ApJS, 127, 59.\newline
Kobayashi,S., \& Sari,R. 2000, ApJ, 542, 819.\newline
Kobayashi,S., Piran,T. \& Sari,R. 1997, ApJ, 490,  92.\newline
Kobayashi,S., Piran,T. \& Sari,R. 1999, ApJ, 513,  669.\newline
Kumar, P. 1999, ApJL, 523, 113.\newline
Kumar, P. \& Piran,T. 2000, ApJ, 535, 152.\newline
Sari,R. 1997, ApJL, 489, L37.\newline
Sari,R 2000 in preparation.\newline
Sari,R. \& Piran,T. 1997, ApJ, 485, 270.\newline
Sari,R., Piran,T. \& Narayan R. 1998, ApJL, 497, L17.\newline
\newpage



\end{document}